\documentclass[preprint,showpacs,preprintnumbers,amsmath,amssymb]{revtex4}

\usepackage{graphicx}% Include figure files
\usepackage{dcolumn}% Align table columns on decimal point
\usepackage{bm}% bold math

\begin{document}

\title
{A balanced homodyne detector for high-rate Gaussian-modulated coherent-state quantum key distribution}

\author{Yue-Meng Chi$^1$, Bing Qi$^1$, Wen Zhu$^1$, Li Qian$^{1,*}$, \\Hoi-Kwong Lo$^1$, Sun-Hyun Youn$^{2,3}$, A. I. Lvovsky$^2$, and Liang Tian$^1$}
\address{$^1$ Center for Quantum Information and
Quantum Control (CQIQC), Dept. of Electrical and Computer
Engineering and Dept. of Physics, University of Toronto, Toronto,
M5S 3G4, Canada}
\address{$^2$ Institute for Quantum Information Science, University of Calgary, Calgary, Alberta T2N 1N4, Canada}
\address{$^3$ Department of Physics, Chonnam National University, Gwangju,
500-757, Korea} \email{l.qian@utoronto.ca}
\begin{abstract}
We discuss excess noise contributions of a practical balanced
homodyne detector in Gaussian-modulated coherent-state (GMCS)
quantum key distribution (QKD). We point out the key generated from
the \emph{original realistic model} of GMCS QKD may not be secure.
In our \emph{refined realistic model}, we take into account excess
noise due to the finite bandwidth of the homodyne detector and the
fluctuation of the local oscillator. A high speed balanced homodyne
detector suitable for GMCS QKD in the telecommunication wavelength
region is built and experimentally tested. The 3 dB bandwidth of the
balanced homodyne detector is found to be 104 MHz and its electronic
noise level is 13 dB below the shot noise at a local oscillator
level of 8.5$\times 10^8$ photon per pulse.

The secure key rate of a GMCS QKD experiment with this homodyne
detector is expected to reach Mbits/s over a few kilometers.
\end{abstract}

\maketitle

%%
%% Copyright 2007, 2008, 2009 Elsevier Ltd
%%
%% This file is part of the 'Elsarticle Bundle'.
%% ---------------------------------------------
%%
%% It may be distributed under the conditions of the LaTeX Project Public
%% License, either version 1.2 of this license or (at your option) any
%% later version.  The latest version of this license is in
%%    http://www.latex-project.org/lppl.txt
%% and version 1.2 or later is part of all distributions of LaTeX
%% version 1999/12/01 or later.
%%
%% The list of all files belonging to the 'Elsarticle Bundle' is
%% given in the file `manifest.txt'.
%%

%% Template article for Elsevier's document class `elsarticle'
%% with numbered style bibliographic references
%% SP 2008/03/01
%%
%%
%%
%% $Id: elsarticle-template-num.tex 4 2009-10-24 08:22:58Z rishi $
%%
%

% \linenumbers

\section{Introduction}
\label{} Quantum key distribution (QKD) based on Gaussian-modulated
coherent-state (GMCS) protocol has attracted a lot of attention
\cite{Grosshans03,Namiki04,Lodewyck05,Legre06,Heid07, QiGMCS07,
Lo08}. Comparing with the BB84 QKD, the GMCS QKD presents several
advantages. The coherent state required by GMCS QKD can be produced
easily by a practical laser source, while the perfect single photon
source required by BB84 QKD is hard to obtain. Although improved
BB84 protocols (such as decoy protocols \cite{Lo05, Ma05, Zhao05,
Zhao06}) are compatible with coherent laser sources, they do require
single photon detectors, which are expensive and have low
efficiency. The homodyne detector in the GMCS QKD, on the other
hand, can be constructed using high efficiency PIN
photodiodes~\cite{Lodewyck05}. The GMCS QKD also has an advantage of
transmitting multiple bits per symbol \cite{Grosshans03,Scarani08}.
The security of the GMCS QKD was first proven against individual
attacks with direct \cite{Grosshans02} or reverse
\cite{Grosshans03,Grosshans04} reconciliation schemes. Security
proofs were then given against general individual attacks
\cite{Grosshans04} and general collective attacks
\cite{Lodewyck07,Navascues06,GarciaPatron06}. To date, three groups
have independently claimed that they have proved the unconditional
security of GMCS QKD \cite{Renner08, Leverrier08, Zhaoyibo08}.

Fiber-based GMCS QKD systems over a practical distance are
challenging and only a few groups have demonstrated QKD experiments
over tens of kilometers\cite{Lodewyck07, QiGMCS07,
QiGMCS08,Fossier08}. Current repetition rates used in those GMCS QKD
experiments are below 1 MHz, which in turn, makes the GMCS QKD less
competitive than the single photon BB84 QKD operating at GHz
repetition rates \cite{Dixon08, Zhang09}. The repetition rate of
GMCS QKD is limited by a few factors: (1) the speed of the homodyne
detector \cite{QiGMCS07}; (2) the speed of the data acquisition
system; and (3) the speed of the classical data processing algorithm
\cite{Lodewyck05}. The speed of date acquisition and classical data
processing can be increased by hardware engineering and are not
fundamental limits in GMCS QKD. In this work, we mostly focus on
increasing the homodyne detector speed and analyzing various excess
noise contributions introduced by a practical homodyne detector.

The balanced homodyne detection used in quantum measurement,
proposed by Yuen and Chan \cite{Yuen83}, plays an important role in
quantum optics \cite{Slusher85,Brei97,Vasi00} and quantum
cryptography \cite{Grosshans03, Lodewyck05, Lodewyck07, QiGMCS07,
Fossier08, Xuan09}. In a balanced homodyne detector (BHD), the
signal to be measured is mixed with a local oscillator (LO) at a
beam splitter. The interference signals from the two output ports of
the beam splitter are sent to two photodiodes followed by a
subtraction operation, and then, amplification may be applied. The
output of a BHD can be made to be proportional to either the
amplitude quadrature or the phase quadrature of the input signal
depending on the relative phase between the signal and the LO. The
output of the BHD can be captured in either frequency \cite{Vasi98}
or time domains \cite{Smithey93, Smithey99, Hansen01, Huisman09}.
For GMCS QKD, measurement in the time domain that is capable of
resolving each individual pulse (representing a weak coherent state)
is required in order to extract random key information
\cite{Grosshans03}. This pulse-resolving requirement demands that
the bandwidth of the detection system be significantly higher than
the repetition rate of the QKD operation, which highlights the
importance of developing high bandwidth BHDs.

In this paper, we develop a broadband BHD suitable for GMCS QKD
operating at a repetition rate of tens of MHz. To predict its
performance in GMCS QKD, we first analyze the excess noise
contributed by this practical BHD. In the GMCS QKD, excess noise is
defined in units of shot noise and includes all noises due to system
imperfections and eavesdropping, which are above and beyond the
vacuum noise associated with channel loss and losses in Bob's
system. It determines the maximum amount of information that could
be obtained by Eve. In the \emph{original realistic model} proposed
in previous GMCS QKD literature \cite{Grosshans03, Lodewyck07,
QiGMCS07}, the excess noise contributed by a BHD is the electronic
noise of the BHD. This model does not consider the excess noise that
originates from other imperfections in a practical BHD and is not
conservative enough in estimating the information possibly be leaked
to Eve. In this paper, we refine the \emph{original realistic model}
which has been widely adopted to calculate key rates for practical
GMCS QKD systems and identify two new noise sources of a practical
homodyne detector: (1) the excess noise caused by the BHD electrical
pulse overlap at the BHD output; and (2) the excess noise caused by
LO fluctuation. Under the \emph{refined realistic model}, we
quantify the various excess noise contributions from the broadband
BHD we constructed. Based on our simulation using the experimentally
determined excess noise of the BHD, secure GMCS QKD key rates using
this BHD is predicted to reach Mbits/s over a few kilometers.

This paper is organized as follows: In Section II, we revisit GMCS
QKD protocol, identify two new excess noise sources, and introduce
the key generation rate formulas based on the \emph{refined
realistic model}. In Section III, we analyze the excess noise
contribution of a practical BHD. In Section IV, we discuss practical
issues in building a high speed BHD, including different temporal
responses of two photodiodes, appropriate pulse duration, and the
BHD linearity and the construction of a high speed HD in GMCS QKD.
In Section V, we will report the performance of the BHD and predict
the key rates by simulation.

\section{Gaussian-modulated coherent-state protocol}
The basic GMCS QKD protocol is as follows: Alice generates two
random sets of continuous variables $x$ and $p$ with a Gaussian
distribution that has a zero average. Alice encodes random bits (key
information) by modulating the amplitude quadrature ($x$) and the
phase quadrature ($p$) of weak coherent states $|{x+ip}\rangle$
(typically less than 100 photons in each pulse) with her
Gaussian-distributed random variable sets $\{x,p\}$. On the
receiver's side, Bob measures either $x$ or $p$ quadrature of the
weak coherent states randomly by using homodyne detection. By
repeating this procedure multiple times, Alice shares a set of
correlated Gaussian variables (called the ``raw key'') with Bob. By
comparing a random sample of their raw key, they can evaluate
parameters of QKD and upper bound on Eve's information. Finally,
they can generate secure key by performing reconciliation.

In the presence of individual attacks, one can estimate the
information leaked to Eve from the amount of excess noise quadrature
noise observed by Bob in excess of standard quantum limit
\cite{Grosshans03}. The most conservative estimation (the
\emph{general model}) assumes all the excess noise is introduced by
eavesdropping, whereas the \emph{original realistic model} assumes
that Eve cannot control the LO or take advantage of the excess noise
generated within Bob's system \cite{Grosshans03}. In the
\emph{original realistic model}, the excess noise has several
contributions: (1) noise due to imperfection outside Bob's system is
denoted as $\varepsilon_A$. This part of noise can be controlled by
Eve; (2) noise from Bob's system that is uncontrollable by Eve,
called $N_{Bob}$. In Refs. \cite{Lodewyck05, Lodewyck07}, the latter
refers to the homodyne detector noise ($N_{hom}$), while in Refs.
\cite{QiGMCS07,QiGMCS08}, it consists both homodyne detector noise
($N_{hom}$) and the noise associated with the photon leakage from
the LO to the signal ($N_{leak}$). In previous papers
\cite{Lodewyck05,Lodewyck07,QiGMCS07, QiGMCS08}, $N_{hom}$ is
regarded to consist of only the electronic noise (i.e. $N_{hom}$ =
$N_{ele}$). In this paper, we refine this \emph{realistic model} and
consider other imperfections of a practical BHD, and conclude that
excess noise caused by a practical BHD ($N_{hom}$) could be divided
into three parts: (1) electronic noise ($N_{ele}$), (2) noise
introduced by electrical pulse overlap due to finite response time
of the BHD ($\varepsilon_{overlap}$) and (3) noise due to local
oscillator fluctuation in the presence of incomplete subtraction of
a BHD ($N_{LO}$).

In Ref. \cite{Lutkenhaus07}, the need to monitor the intensity of
the LO for security proofs in discrete QKD protocol embedded in
continuous variables has been discussed. In GMCS QKD experiment,
Alice and Bob can monitor LO, and discard pulses with large
intensity changes in LO. However, there is always a small
measurement error due to imperfect measurement instrument.
Therefore, it is reasonable to assume there is a small amount of LO
fluctuation that Eve can take advantage of. Therefore, in this
\emph{refined realistic model}, $\varepsilon_{overlap}$ and $N_{LO}$
generated by a BHD, as well as $N_{leak}$ associated with leakage LO
photons, are all considered controllable by Eve. $N_{LO}$ is caused
by imperfect subtraction of BHD in the presence of LO intensity
fluctuation while $N_{leak}$ is due to the interference between
leakage photons and LO photons.

%\cite{Grosshans03, Lodewyck05, Scarani08,
%Grosshans04}.
%Alice's modulation variance is $V_A$ (variance of $x$ or $p$
%quadrature modulated by Alice), and $V = V_{A}+1$ is the quadrature
%variance of the coherent state prepared by Alice (1 is the shot
%noise of a coherent state). The channel efficiency (transmission) is
%$G$, and the total efficiency of Bob's device (optical loss and
%detector efficiency) is $\eta$. $\chi$ is the equivalent noise
%measured at the input, which is composed of quantum noise of channel
%$\chi_{vac}$ and excess noise $\epsilon$.
%, noise that can be controlled by Eve
%$\epsilon_E$ and noise which is from Bob's devices and cannot controled by Eve %$N_{Bob}$.
Following an approach similar to that in \cite{Grosshans03}, we will
now present the GMCS QKD key rate formulas based on \emph{refined
realistic model}. The mutual information between Alice and Bob
$I_{AB}$ is determined by the Shannon entropy \cite{Shannon48}.
According to Refs. \cite{Grosshans03, Lodewyck05},
%\frac{1}{2}\log_2 \frac{V_B}{V_{B|A}}
\begin{equation}\label{eq:IAB}
I_{AB} = \frac{1}{2}\log_2 [(V+\chi)/(1+\chi)],
\end{equation}
where
\begin{equation}\label{eq:chi}
\chi = \chi_{vac}+\varepsilon = \frac{1-\eta G}{\eta G}+\varepsilon.
\end{equation}
In Eq. (\ref{eq:IAB}), $V = V_{A}+1$ is the quadrature variance of
the coherent state prepared by Alice (1 is the shot noise of a
coherent state) and $V_A$ is Alice's modulation variance (variance
of $x$ or $p$ quadrature modulated by Alice). In Eqs. (\ref{eq:IAB})
and (\ref{eq:chi}), $\chi$ is the equivalent noise measured at the
input, which is composed of ``vacuum noise'' $\chi_{vac}$ (noise
associated with the channel loss and detection efficiency of Bob's
system) and ``excess noise'' $\varepsilon$ (noise due to the
imperfections in a non-ideal QKD system). $G$ is the channel
efficiency (transmission), and $\eta$ is the total efficiency of
Bob's device (optical loss and detector efficiency).

%There are two models that can be used to estimate Eve's information.
%The ``general model'' assumes that losses and noise in Bob's system
%can be controlled by Eve \cite{Grosshans03}. In contrast, in a
%``realistic model'', we assume that Eve has no control over Bob's
%system.

%Under the ``general model'', the mutual information between Bob and
%Eve $I_{BE}$ is
%\begin{equation}\label{eq:IBE}
%I_{BE} =\frac{1}{2}\log_2 [(\eta G)^2(V+\chi)(V^{-1}+\chi)].
%\end{equation}

\begin{figure}[!hb]\center
\includegraphics[width=10cm]{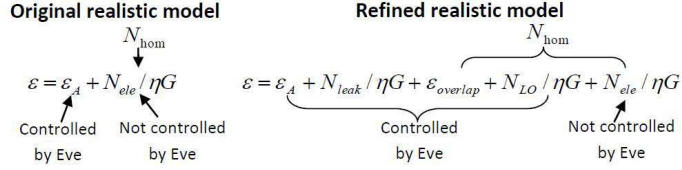}
\caption{Various noise terms in the \emph{original realistic model}
and the \emph{refined realistic model}.}\label{noisemodel}
\end{figure}

We will now discuss the key rate formulas for the case of the
\emph{refined realistic model}, which we defined earlier in this
Section. Under the \emph{refined realistic model}, noise that can in
principle be controlled by Eve ($\varepsilon_{E}$) includes (1)
$\varepsilon_A$ due to imperfections outside Bob's system; (2)
$\varepsilon_{overlap}$ introduced by electrical pulse overlap due
to finite response time of the BHD; (3) $N_{LO}$ due to LO
fluctuations in the presence of incomplete subtraction of a BHD and
(4) $N_{leak}$ associated with the leakage from LO to signal.
%, which is
%considered to be the only part of excess noise controlled by Eve
%under the realistic model in previous literatures;
Excess noise that is out of Eve's control ($N_{Bob}$) is electronic
noise from the homodyne detector ($N_{ele}$). Therefore, the total
excess noise $\epsilon$ can be written as \cite{Grosshans03}
\begin{equation}\label{eq:excessnoise}
\varepsilon = \varepsilon_E + N_{Bob}/\eta G,
\end{equation}
where $\varepsilon_{E}=
\varepsilon_{A}+\varepsilon_{overlap}+N_{LO}/\eta G+N_{leak}/\eta G$
and $N_{Bob} = N_{ele}$. $\varepsilon_A$ and $\varepsilon_{overlap}$
are referring to the input. $N_{LO}$, $N_{leak}$ and $N_{ele}$ are
defined from the output and need to be divided by $\eta G$ when we
convert them to the input. Figure \ref{noisemodel} summarizes the
various noise terms considered in the \emph{original realistic
model} and the \emph{refined realistic model}. $N_{leak}$ is mostly
determined by the design of the QKD system rather than by the BHD.
Since our main goal is to study the excess noises contributed by the
BHD, we simply assume $N_{leak}=0$ in this paper.

From Eqs. (\ref{eq:chi}) and (\ref{eq:excessnoise}), the equivalent
input noise is
\begin{equation}\label{eq:chi2}
\chi = \frac{1-\eta G}{\eta G}+\varepsilon_E +\frac{N_{Bob}}{\eta
G}.
\end{equation}

%Bob's quadrature variance is given by $V_B=\eta G(V+\chi)$, while
%the conditional variance under the "realistic model" is
%\begin{equation}\label{eq:VBgivenE}
%V_{B|E} = \frac{\eta}{1-G+G(\epsilon_A+V^{-1})}+(1-\eta)+N_{Bob}.
%\end{equation}

With a reverse reconciliation scheme, the mutual information shared
by Bob and Eve under the \emph{refined realistic model} is
\begin{equation}\label{eq:IBE2}
I_{BE}=\frac{1}{2}\log_2[\frac{\eta GV_A +1+\eta G
\varepsilon}{\eta/{(1-G+G\varepsilon_E+GV^{-1})}+1-\eta+N_{Bob}}].
\end{equation}

If a reverse reconciliation algorithm \cite{Grosshans03} is adopted,
the secure key rate is
\begin{equation}\label{eq:MutualInfoGeneral}
\Delta I = \beta I_{AB} - I_{BE}.
\end{equation}
where $\beta$ is the reconciliation efficiency ($\beta \leq$ 1). In
real QKD systems, $\beta$ is 0.9 in Ref. \cite{Fossier08} and 0.898
in Ref. \cite{Lodewyck07}. If the laser repetition rate of QKD
experiment is $R$ Hz, the secure key per second can be written by
\begin{equation}\label{eq:Info}
\Delta I_{second} = (\beta I_{AB} - I_{BE})\times R.
\end{equation}

%The secure key is calculated by Eq. (\ref{eq:MutualInfoGeneral})
%given a reconciliation efficiency $\beta$.

\section{Excess noise contributed by the BHD in a GMCS QKD}
As previously stated, excess noise represents the amount of
information that could possibly be leaked to Eve in a GMCS QKD
system and is important in estimating the amount of secure
information.

In this section, we will evaluate various sources of the excess
noise for a practical BHD.

\subsection{BHD electronic noise}
Electronic noise $N_{ele}$ of a BHD is mainly contributed by thermal
noise of electronic components and amplifier noise \cite{Bachor04}.
Since shot noise scales with LO power and electronic noise is
independent of the LO power \cite{Lvovsky08}, by measuring the BHD
noise as a function of the LO power when vacuum is sent to the
signal port, we can quantify the electronic noise in units of shot
noise. Electronics noise in a BHD has been discussed in
\cite{Appel07}.

\subsection{Excess noise due to electrical pulse overlap}
Ideally, the secure key rate of a GMCS QKD system is proportional to
its operation rate. However, in practice, the BHD has a finite
bandwidth. As the laser pulse repetition rate approaches the
bandwidth of the BHD, we will expect a non-negligible overlap
between adjacent electrical pulses at the output of the BHD. If the
electrical pulses have overlap in the time domain, the measured
quadrature value contains contributions from adjacent pulses.

We will estimate the amount of excess noise contributed by the
electrical pulse overlap. The exact relation between the electrical
pulse width $\tau$ and the BHD bandwidth $B$ depends on the
electrical pulse shape. We have experimentally found that the
relation $\tau \sim 1/B$ applicable to our homodyne detector.
%For a BHD bandwidth of $B$, its output
%electrical pulse width is $\sim 1/(2\pi B)$. If we assume the electrical
%pulse has a Gaussian shape in time, then
In this case, we can estimate the overlap by writing the following
functions for two consecutive pulses: (a) $e^{-(t-1/R)^2/2\tau ^2}$
and (b) $e^{-t^2/2\tau ^2}$, where $R$ is the laser repetition rate
and $\tau$ is the Gaussian pulse width. If the quadrature value is
determined by the peak of the measured electrical pulse, the
contribution of pulse (a) to pulse (b) is $e^{-\frac{B^2}{2R^2}}$.
Since each pulse has two adjacent pulses, the excess noise
contributed by electrical pulses overlap (referring to the input) is
\begin{equation}
\varepsilon_{overlap} = 2 V \times (e^{-\frac{B^2}{2 R^2}})^2 = 2
(V_A+1) \times e^{-\frac{B^2}{R^2}}.
\end{equation}
where $V_A$ is Alice's modulation. We remark that the excess noise
due to the overlapping between adjacent pulses could be further
reduced by deconvolution \cite{Huisman09}.

\begin{figure}[!hb]\center
\includegraphics[width=10cm]{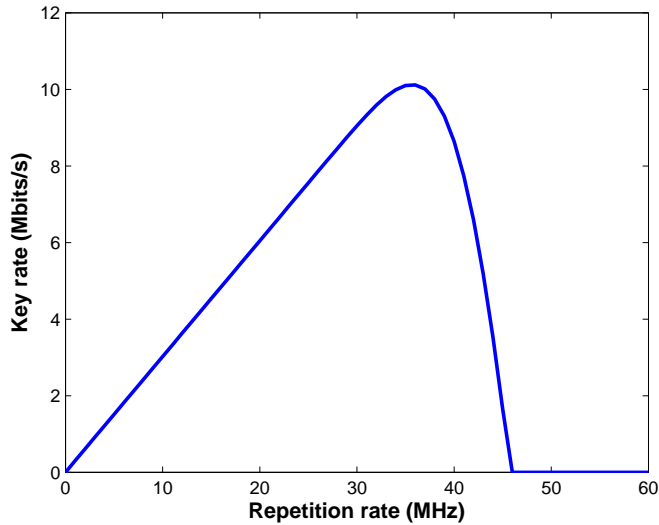}
\caption{GMCS QKD secure key generation rate as a function of the
laser repetition rate under the \emph{refined realistic model}. The
bandwidth of the BHD is 100 MHz. The simulation parameters are from
Ref. \cite{QiGMCS07}, $V_A = 16.9$, $G = 0.758$, $\eta = 0.44$,
$\varepsilon_A$ = 0.056, $N_{ele} = 0.045$, and $\beta$=0.898. In
this simulation, $N_{LO}$ = $N_{leak}$ = 0.}\label{RR}
\end{figure}

By decreasing this repetition rate, we can reduce the excess noise
caused by overlap. However, the GMCS QKD key rate per second will be
reduced too. In Fig. \ref{RR}, we simulate the GMCS QKD key rate per
second as a function of the repetition rate using Eqs.
(\ref{eq:IAB}), (\ref{eq:IBE2}), (\ref{eq:MutualInfoGeneral}) and
(\ref{eq:Info}). With a BHD bandwidth of 100 MHz, the optimal pulse
repetition rate is around 36 MHz. At repetition rates beyond $\sim$
46 MHz, there will not be any secure key generated. At low
repetition rate, the excess noise due to electrical pulse overlap is
negligible compared to other excess noise contribution
($\varepsilon_A$), and the key rate per second is almost
proportional to the repetition rate. As the repetition rate is
increased beyond a critical point, the excess noise due to overlap
is dominant and the key rate drops quickly with the repetition rate.

\subsection{Excess noise contributed by LO
fluctuations}\label{noise:LO} One of the advantages of BHD is that
ideally the fluctuations of LO will be canceled after the
subtraction. However, in a practical BHD, the positive and negative
pulses cannot be canceled completely due to several reasons, such as
different quantum efficiencies of the two photodiodes, different
temporal responses of the photodiodes and the subsequent electronic
amplifiers, or different optical intensities of the two balanced
beams. The difference can be partially compensated, for example, by
adjusting the losses and the relative delay of the two balanced
arms, however, it cannot be completely canceled out. The remaining
difference also varies with LO power. The consequence is that the
fluctuation of the LO power will contribute to the excess noise.

The quadrature measurement corresponds to the time integrated
electronic response of the detector. Neglecting the shot noise, this
response equals
\begin{equation}\label{balance}
 N_{LO} = I_{LO} [G_1 t^2 -G_2 r^2] ,
\end{equation}
where  $I_{LO}$ is the number of photons in the local oscillator
pulse, $t$ is the beam splitter transmittance, $r$ is reflectivity
and $G_{1,2}$ are the time integrated gains of the amplifiers
associated with the two photodiodes. We assumed that the quantum
efficiency of the photodiodes is 1 and the signal is in the vacuum
state. On the other hand, given that the variance in the number of
photoelectrons in each photodiode due to the shot noise equals to
the number of incident photons, we obtain the the output shot noise
as
\begin{equation}\label{shot}
 \langle N_{shot}^2 \rangle=I_{LO} [G_1^2 t^2+G_2^2 r^2]  .
\end{equation}
If the relative fluctuation of the LO power is $\sqrt{\langle \Delta
I_{LO}^2 \rangle}/I_{LO}= f$, the mean square fluctuation in the
number of output photoelectrons in the units of shot noise is
\cite{Raymer95}
\begin{equation}\label{balancetoshot}
 \frac {\langle \Delta N_{LO}^2 \rangle} {\langle N_{shot}^2 \rangle}  = I_{LO} f^2 \delta^2 \quad  \textrm{with} \quad \delta= \frac {G_1 t^2 -G_2 r^2} {\sqrt{G_1^2 t^2+G_2^2 r^2}} .
\end{equation}
For a well-balanced detector, $t^2\approx r^2\approx 1/2$ and
$G_1\approx G_2$. In this case, the above expression can be written
as $\delta\approx\delta_{opt}+\delta_{el}$, where $\delta_{opt}=t^2
-r^2$ is the imbalance of the optical beam splitter whereas
$\delta_{el}=(G_1-G_2)/(G_1+G_2)$ is the electronic characteristic
of the balanced detector related to its common-mode rejection ratio
(CMRR). In what follows, it is convenient to discuss $N_{LO}$ in
terms of generalized CMRR which is measured in decibels and defined
as
\begin{equation}\label{CMRREq}
\rm {CMRR} = -20 \log_{10} (2\delta).
\end{equation}

The magnitude of $N_{LO}$ can be estimated from the Taylor
decomposition of the noise variance as a function of the local
oscillator power. The shot noise variance is proportional to the LO
level, whereas $N_{LO}$ depends on it quadratically \cite{Bachor04}.
We note that for determining $N_{LO}$, only time-integrated response
functions (over the bandwidth of the homodyne detector) of the
photodetector-amplifier systems are relevant; Faster-varying
differences in the time dependent shapes of these functions play no
role.

\begin{figure}[!t]\center
\includegraphics[width=10cm]{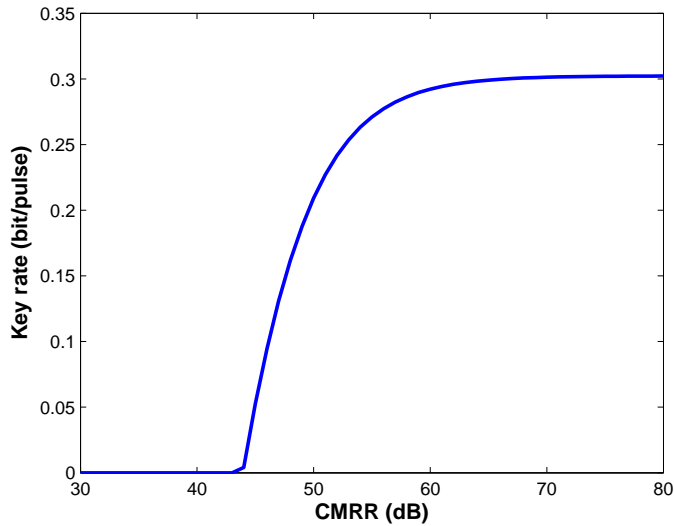}
\caption{GMCS QKD secure key rate per pulse as a function of the
CMRR using the \emph{refined realistic model}. Here, we assume
$10^8$ LO photons/pulse and 1 $\%$ LO fluctuation. The simulation
parameters are from Ref. \cite{QiGMCS07}, $V_A = 16.9$, $G = 0.758$,
$\eta = 0.44$, $\varepsilon_A$ = 0.056, $N_{ele} = 0.045$, and
$\beta$=0.898. In this simulation, $\varepsilon_{overlap}$ =
$N_{leak}$ = 0. }\label{CMRR}
\end{figure}
With the same GMCS QKD parameters used to produce Fig. \ref{RR}, we
simulate the GMCS QKD secure key rate as a function of the CMRR of
the BHD in Fig. \ref{CMRR}. When CMRR is lower than 55 dB (where key
rate is 90 $\%$ of the maximum), key rate drops quickly as the CMRR
drops. To obtain a positive key rate, the CMRR of the BHD should be
greater than $\sim 44$ dB. When CMRR is greater than 55 dB, the
secure key rate will not improve too much by increasing the CMRR
since other excess noise contribution ($\varepsilon_A$) is dominant.
\section{Construction and performance}
\begin{figure}[!hb]\center
\includegraphics[width=10cm]{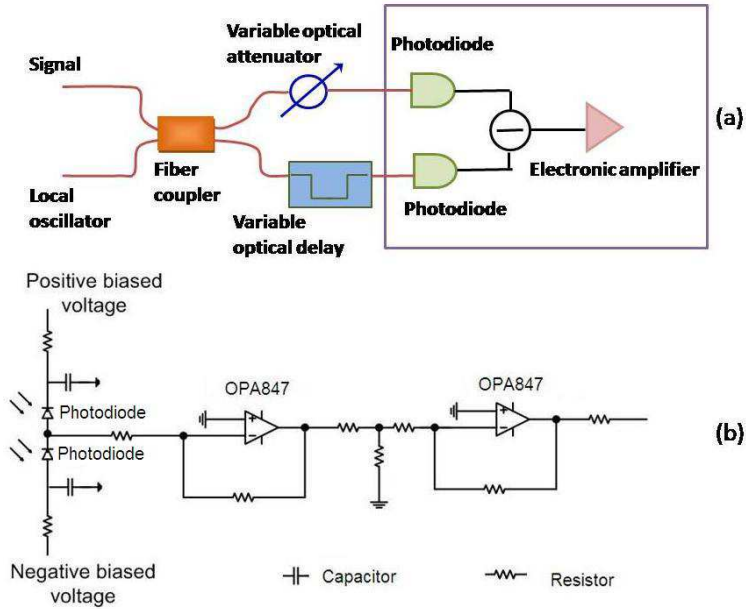}
\caption{(a) Balanced homodyne detector (BHD) schematic in the
telecommunication wavelength. The red lines are optical paths and
the black lines are electrical cables. (b) Simplified BHD electronic
circuit (the components in the right square of (a)).}\label{HDsetup}
\end{figure}

In this section, we will present our construction and test results
of a high speed BHD in the telecommunication wavelength region. We
will also predict the excess noise and secure key rate by using this
BHD in a GMCS QKD experiment.

\subsection{Schematic}
Figure \ref{HDsetup} (a) shows a schematic of our balanced homodyne
detection system. In the telecommunication wavelength region, the
signal and the LO beams will interfere at a two-by-two fiber coupler
 with a splitting ratio of 50:50. A
variable optical attenuator and a variable optical delay are placed
in the output paths of the fiber coupler, for adjusting losses and
the lengths of the two paths accurately. Two photodiodes will detect
the interference beams of the signal and the LO after precise
balancing of time and intensity. Finally, a subtraction of the
photocurrents generated by the two photodiodes is performed and the
differential signal is amplified. To avoid disturbances from the
environment, we used an enclosure to isolate the system of Fig.
\ref{HDsetup} (a).

In the electronic circuit shown in Fig. \ref{HDsetup} (b), two
InGaAs photodiodes from Thorlabs (FGA04, 2 GHz bandwidth, quantum
efficiencies:  90 $\%$ and $93 \%$) are reversely biased. The
differential signal is amplified by two OPA847 operational
amplifiers. The whole BHD circuit is built on a custom-designed
printed circuit board. To minimize the parasitic capacitance, two
photodiodes with short electrical contact legs are placed very close
to each other.
\subsection{Linearity}
\begin{figure}[!hb]\center
\includegraphics[width=12cm]{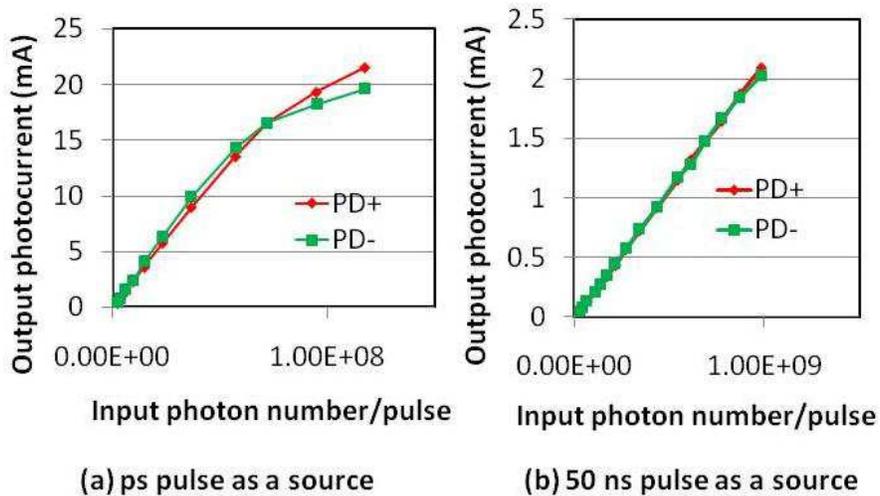}
\caption{Photodiode linearity test. The peak photocurrent as a
function of the input photon number in each pulse when (a) $\sim$ 1
ps width laser as a source; (b) 50 ns width laser as a
source}\label{PDlinearity}
\end{figure}

In GMCS QKD, continuous Gaussian random numbers encoded on each
pulse have to be recovered by the balanced homodyne detection on
Bob's side. To ensure the BHD output is proportional to the electric
field quadrature of each pulse, the linearity of the BHD has to be
guaranteed. In practice, the photodiode and electronic amplifiers
can both have nonlinearities. A proper pulse width should be
carefully chosen to guarantee that the photodiodes are working in
their linear regions. In the test of the photodiode linearity, we
send pulsed light to only one photodiode while blocking the other
one. At a laser repetition rate of 10 MHz, we measure the output
photocurrent generated by the photodiode (before it goes to the
electronic amplifiers) at different incident optical powers using an
oscilloscope. In Fig. \ref{PDlinearity}, we compare the output
electrical pulse peak current when a laser source with (a) $\sim$1
ps or (b) 50 ns pulse duration is used. We can see from Fig.
\ref{PDlinearity} (a), the photodiodes saturate at a low optical
input photon number per pulse than that of (b). In fact, the high
peak power of the $\sim$1 ps-pulse ($\sim$ 18 W) saturates the
photodiodes. In the case of 50-ns pulse as a source (Fig.
\ref{PDlinearity} (b)), photodiodes are working in their linear
regions (4 $\%$ deviation) up to $10^9$ photons/pulse.

\begin{figure}[!hb]\center
\includegraphics[width=12cm]{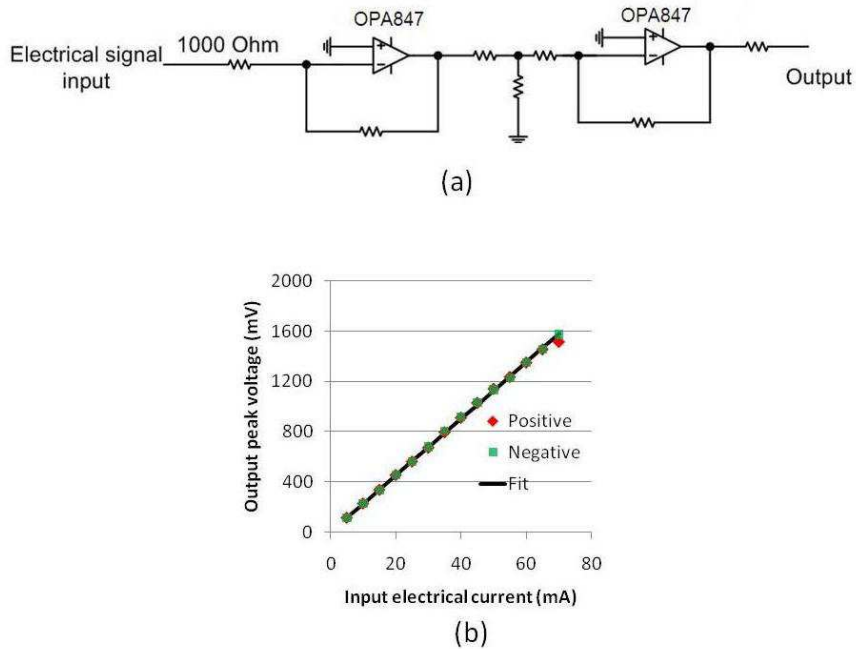}
\caption{(a) Electronic amplifier linearity test circuit; (b) Output
electrical pulse peak voltage as a function of input electrical
current.}\label{linearity2}
\end{figure}

The linearity test of the electronic amplifiers is shown in Fig.
\ref{linearity2}. By sending positive or negative electrical pulses
(50-ns width, 10 MHz repetition rate) to the electrical amplifiers
shown in Fig. \ref{linearity2} (a), we measure the output electrical
pulse peak voltage as a function of the input electrical current.
The trans-impedance gain is measured to be 22 kV/A in Fig.
\ref{linearity2} (b). The trans-impedance gains for the positive and
negative pulses are almost equal with less than 1 $\%$ deviation
from their linear fits.

\subsection{BHD bandwidth}
\begin{figure}[!hb]\center
\includegraphics[width=10cm]{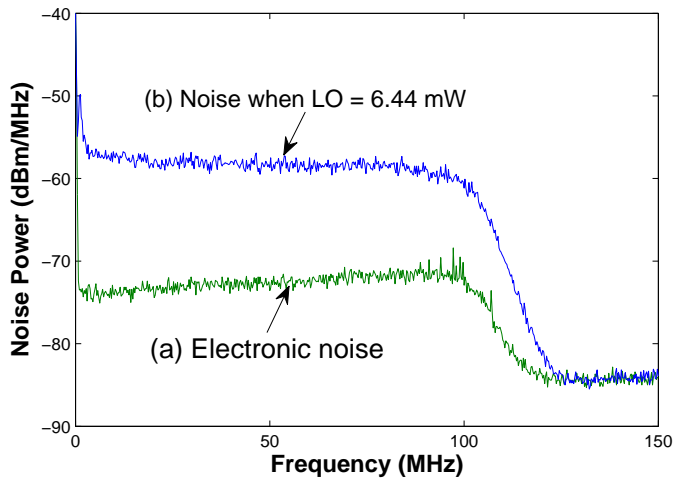}
\caption{(a) Electronic noise (b) BHD noise at an LO of 6.44
mW.}\label{cwfreq}
\end{figure}
We first characterize the bandwidth of our BHD by sending a CW LO.
In this case, the residual signal caused by different temporal
responses of the photodiodes can be eliminated by adjusting the loss
in one arm (Fig. \ref{HDsetup} a). Using an RF spectrum analyzer,
the spectral noise is measured and shown in Fig. \ref{cwfreq}. In
the frequency domain, the trace (a) is the electronic noise and is
measured when no optical signal is sent to the BHD. We can see the
3-dB bandwidth of the BHD is 104 MHz. Trace (b) is measured when
6.64 mW CW LO is sent to the BHD. The noise includes electronic
noise and shot noise.

\subsection{Homodyne detector noise measurement in the time
domain}\label{Chap3:shotnoise}

In GMCS QKD, each pulse will be measured individually. In the time
domain, we first performed HD noise measurement at a pulse
repetition rate of 10 MHz and obtained 12 dB shot noise to
electronic noise ratio at an LO photon level of 8.2 $\times 10^8$.
We further increase the repetition rate to 32 MHz and will
demonstrate our results here.

\begin{figure}[!hb]\center
\includegraphics[width=10cm]{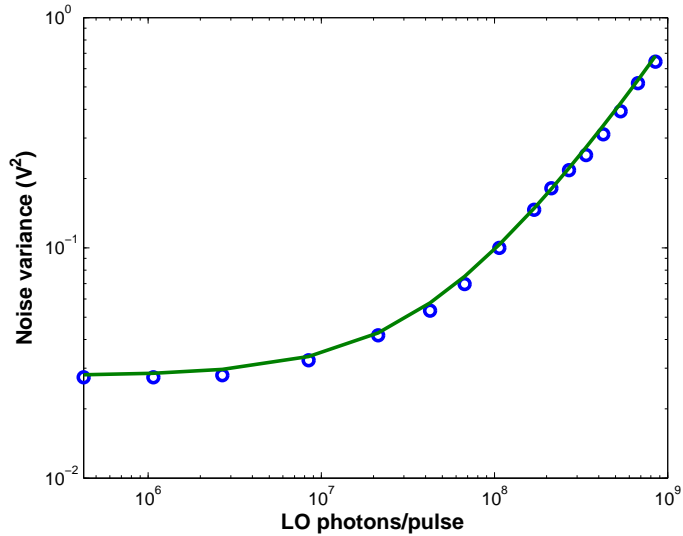}
\caption{Total noise of BHD as a function of the LO photon number
per pulse.}\label{noisetime2}
\end{figure}

With a 16-ns-width pulsed LO (5-ns edge time) at a repetition rate
of 32 MHz, the total noise of BHD of each pulse is obtained by
integrating the BHD output voltage over the pulse region. With an
oscilloscope sampling rate of 20 G samples/s, and an integration
time window of 20 ns in each cycle, each pulse quadrature is
obtained from 400 sample points. Noise variance is obtained from 640
pulses. Fig. \ref{noisetime2} shows the BHD noise variance as a
function of the LO photon number per pulse. The measured homodyne
detector noise includes: (1) electronics noise $N_{ele}$, (2) shot
noise, and (3) noise associated with LO fluctuation $N_{LO}$.
Because Fig. \ref{noisetime2} displays the \emph{square} variance,
the shot noise should appear linear to the LO level, and $N_{LO}$,
which is linear to LO level in shot-noise units becomes
quadratically dependent on LO level when plotted in V$^2$ units.
Note that $\varepsilon_{overlap}$ is neglected since it is much less
than the shot noise when the signal is vacuum. We distinguish noises
by separating the quadratic LO-dependent ($N_{LO}$), the linear
LO-dependent (shot noise) and LO-independent ($N_{ele}$) components
of the BHD output signal. From the experimental results, the total
variance of the BHD output signal (in V$^2$)can be written as $y
=8.0\times 10^{-20}\cdot I_{LO}^2+7.0\times 10^{-10} \cdot I_{LO} +
0.028$ where $I_{LO}$ is the LO photon number per pulse. The
coefficient of determination is 0.999 \cite{Notefitting}. The
electronic noise $N_{ele}$ (in shot noise unit) can be determined
from the ratio of the third term and the second term, which is
4.0$\times 10^{7}/I_{LO}$. We find the shot noise to electronic
noise ratio is 13 dB at an LO photon level of $8.5\times 10^8$ per
pulse. In the meantime, $N_{LO}$ (in shot noise unit) can be
determined from the ratio of the first term to the second term,
which is $1.1 \times 10^{-10}\times I_{LO}$.

As a simple check of the randomness of the noise, we measure the
correlation coefficient (CC) between adjacent sampling results. CC
is defined as
\begin{equation}
{\rm CC} =
\frac{{E(X(n)X(n+1))}-E(X(n))E(X(n+1))}{\sqrt{E(X(n)^2)-E^2(X(n))}\sqrt{E(X(n+1)^2)-E^2(X(n+1))}}.
\end{equation}
while $X(n)$ is the quadrature value of the $n$th pulse. At
3.4$\times 10^8$ LO photons/pulse, the correlation coefficient
between consecutive pulses is 0.051, which is comparable with other
BHDs reported in Ref. \cite{Okubo08} (0.04) and Ref.
\cite{Haderka09} (0.07). We can use the CC to determine the upper
bound of the excess noise caused by electrical pulse overlap
$\varepsilon_{overlap}$. In GMCS QKD, with the quadrature variance
of the coherent state prepared by Alice $V$, the excess noise due to
the overlap between pulses will be $V \times  {\rm CC}^2
=(V_A+1)\times \rm {CC}^2$ (referring to the input) \cite{Explain4}.
Assuming Alice's modulation $V_A$ = 16.9 \cite{QiGMCS07} and each
pulse has two neighboring pulses, we derive the excess noise caused
by BHD pulse overlap to be 0.044 referring to the input.

\subsection{Common mode rejection ratio}\label{Chap3:CMRR}
\begin{figure}[!hbt]\center
  \includegraphics[width=10cm]{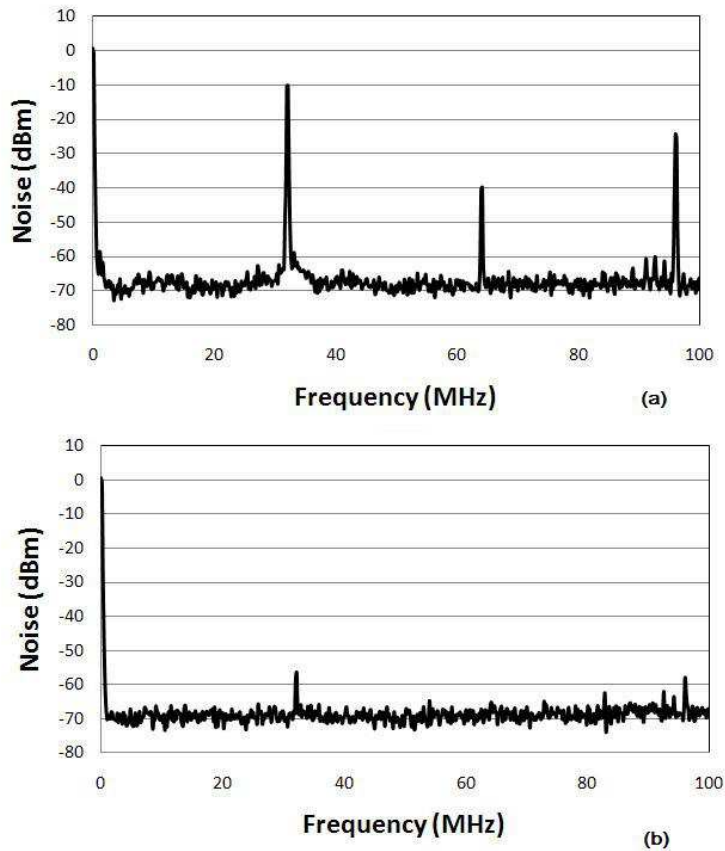}
  \caption{Noise spectrum at an LO power of 24.6 $\mu$W
  when (a)two photodiodes are illuminated; (b)one photodiode is blocked. Resolution bandwidth: 100 kHz}
  \label{CMRR2}
  \end{figure}

To quantify the subtraction capability of the BHD, we measure the
CMRR. In the frequency domain, we obtain CMRR by measuring the
spectral power difference at the repetition rate of 32 MHz in two
cases: (a) one photodiode is blocked and the other is illuminated
(b) both photodiodes are illuminated. At an LO power of 24.6 $\mu$W,
the spectral noise for both cases is shown in Fig. \ref{CMRR2}. The
CMRR is obtained to be 46.0 dB.

\subsection{Excess noise evaluation and key rate simulation for a GMCS QKD experiment}
\begin{table}[!b]
  \center
  \caption{Excess noise contributions by the BHD (in the shot noise unit). $I_{LO}$ indicates the LO photon number per pulse.}\label{table:noises}
    \begin{tabular}{c c c c c}
      \hline
        & Referring to the input & & & Referring to the output \\
      $N_{ele}$ & $4.0\times 10^{7}/(\eta G I_{LO})$ & & & $4.0\times 10^{7}/I_{LO}$ \\
      $\varepsilon_{overlap}$ & 0.044 & &  & 0.044$\times \eta G$ \\
      $N_{LO}$ & $1.1\times 10^{-10}\cdot I_{LO}/\eta G$ &  & & $1.1\times 10^{-10}\cdot I_{LO}$\\
      \hline
    \end{tabular}
\end{table}

\begin{figure}[!hb]\center
\includegraphics[width=10cm]{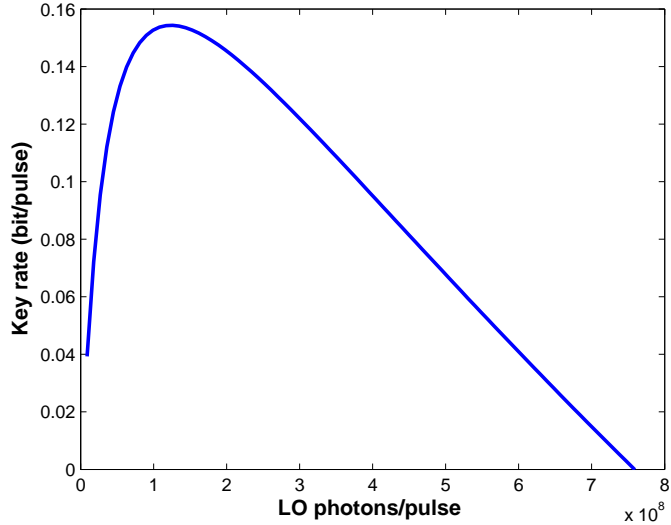}
\caption{Optimization of LO photon number under the \emph{refined
realistic model}. The simulation parameters are from Table
\ref{table:noises} and Ref. \cite{QiGMCS07}, $G = 0.758$, $V_A =
16.9$, $\eta = 0.44$, $\varepsilon_A$ = 0.056, and
$\beta$=0.898.}\label{keyratevsLO}
\end{figure}

 Under this \emph{refined
realistic model}, we identify new excess noise sources of a
practical BHD. Various sources of excess noise contributed by this
BHD are summarized in Table \ref{table:noises}. Given this practical
BHD, we can also optimize operation parameters based on the
\emph{refined realistic model}. In Fig. \ref{keyratevsLO}, we
simulate the key rate per pulse as a function of the LO level. The
key rate under the \emph{refined realistic model} will reach the
maximum at an LO photon number of $1.3\times 10^8$ per pulse,
because there is a tradeoff between $N_{LO}$ (increasing with LO
level) and $N_{ele}$ (decreasing with LO level).

%We can see that the general model which assumes all
%excess noise can be controlled by Eve is the most conservative
%model. The key rate under the \emph{general model} is also the
%lowest among the three models.

%The old ``realistic model'' considers the noise of a BHD is only
%from the electronic noise and cannot be controlled by Eve. However,
%it actually underestimates the possible leaked information to Eve
%when a practical BHD is used in the QKD system. In GMCS QKD, for
%example, Eve could potentially pass both signal and LO through a
%highly dispersive fiber (but not lossy) to increase the pulse
%duration. Eve can also change the LO fluctuation to increase the
%excess noise $\epsilon_{LO}$ thus obtain information. Therefore, in
%principle, Bob should monitor the LO fluctuation and the pulse
%width, and use the model suggested here to take into account the
%resulting excess noise.

\begin{figure}[!th]\center
\includegraphics[width=10cm]{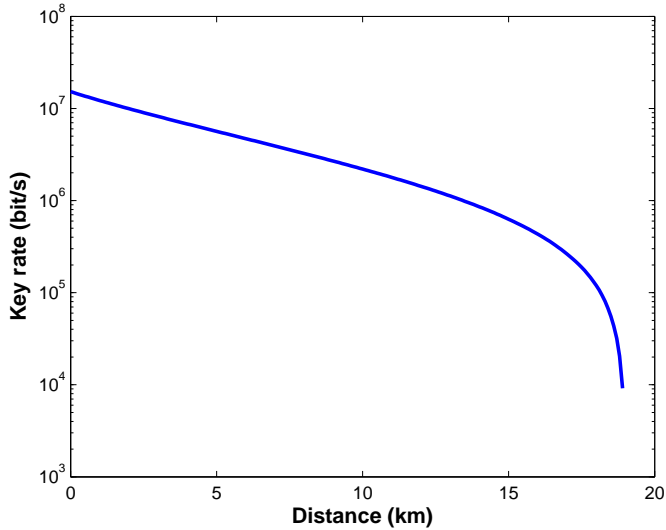}
\caption{QKD secure key rate under the \emph{refined realistic
model} as a function of the transmission distance when the
repetition rate is 32 MHz based on the performance of our BHD. The
simulation parameters are from Table \ref{table:noises} and Ref.
\cite{QiGMCS07}, $V_A = 16.9$, $\eta = 0.44$, $\varepsilon_A$ =
0.056, and $\beta$=0.898. Fiber loss is 0.21 dB/km. For each
distance, the LO level is chosen to maximize the secure key rate. No
secure key rate can be generated beyond 20 km due to the excess
noise.}\label{keyrate}
\end{figure}

In Fig. \ref{keyrate}, we simulate the secure key rate of GMCS QKD
using this BHD under the \emph{refined realistic model} by choosing
the optimal LO level for each transmittance. With this high speed
BHD allowing a repetition rate of tens of MHz, the secure key
generation rate of GMCS QKD can be improved by 1-2 orders of
magnitude comparing to current systems at 500 kHz repetition rate in
Ref. \cite{Lodewyck07} (with a key rate of 2 kbits/s over 25 km
fiber), and in Ref. \cite{Fossier08} (with a key rate of 8 kbits/s
over 3 dB loss channel) and at 100 kHz in Ref. \cite{QiGMCS08} (with
a key rate of 5 kbits/s over 20 km fiber). From the key rate
simulation in Fig. \ref{keyrate}, we expect to achieve a few Mbits/s
over a short distance in future GMCS QKD under the \emph{refined
realistic model}.

\section{Conclusion}\label{con}

In conclusion, we have analyzed the excess noise contributed by a
practical BHD and refined the \emph{realistic model}. The electronic
noise $N_{ele}$, excess noise due to electrical pulse overlap
$\varepsilon_{overlap}$ and excess noise caused by LO fluctuations
in the presence of incomplete subtraction $N_{LO}$ are three excess
noise sources for a practical BHD. They introduce a security
loophole since Eve can monitor the pulse width and slightly change
the LO intensity. Implementing attacks with current technology to
GMCS QKD will be an interesting research direction to explore.

We also developed a high speed BHD with a 104 MHz bandwidth in the
telecommunication wavelength region for the first time. A comparison
of the specifications between our BHD and other high speed BHD is
shown in Table \ref{compare2}. We achieved a
shot-noise-to-electronic-noise ratio of 13 dB in the time domain at
a pulse repetition rate of 32 MHz. The BHD has a high CMRR of 46.0
dB. Various sources of excess noise introduced by this practical BHD
are identified, and their contributions to excess noise are
evaluated. With this BHD, the key generation rate of GMCS QKD
experiments is expected to reach a few Mbits/s under the
\emph{refined realistic model}.
\begin{table}[!b]
  \center
  \caption{A comparison between high speed BHD}\label{compare2}
    \begin{tabular}{c c c c c}
      \hline
        &  \cite{Okubo08} & \cite{Haderka09}  & \cite{Zavatta02} & Our BHD \\
      Wavelength (nm) &  1064 & 800 & 786 & 1550 \\
      Bandwidth (MHz)  & $\sim$ 250 & $\sim$ 70 & $\geq$ 82 & $\sim$ 100 \\
      CMRR (dB) & 45& 61.8 & 42 & 46.0\\
      Shot-noise-to-electronic-noise ratio (dB) & 7.5 & 12 & - & 13\\
      \hline
    \end{tabular}
\end{table}

%By
%verifying the linearities of the photodiodes, testing the linearity
%and gain of electronic amplifiers and choosing a proper pulse width,

\section{Acknowledgement}
We thank CFI, CIPI, the CRC program, CIFAR, MITACS, NSERC, OIT, and
QuantumWorks for their financial support.



%% References







% \end{thebibliography}


\begin{thebibliography}{00}
\bibitem{Notefitting} If the measured HD noise is represented by
$y'(i)$ and the fitting noise is represented by $y(i)$ while $i$ is
the index for each LO level, the coefficient of determination is
determined by $1-\frac{\sum_i [y(i)'-y(i)]^2}{\sum_i
[y'(i)-\bar{y}]}$.
\bibitem{Explain4}
%For two sequences $X$ and $Y$, the correlation coefficient (CC) is $\frac{{E(XY)}-%E(X)E(Y)}{\sqrt{E(X^2)-E^2(X)}\sqrt{E(Y^2)-E^2(Y)}}$. Considering $X$ %and $Y$ having one delay pulse, if we write $Y$ into $X$ dependent and independent %components,i.e. $Y=X'+aX$ ($X'$ is independent of $X$, $a$ is a constant), the CC will %be $\frac{{E(XY)}}{E(X^2)} = a$, where we assume $E(X)=E(Y)=0$. In our measurement, %the CC = 0.034. The overlap factor between consecutive pulse is 0.034. With a GMCS QKD %modulation of 10, the estimated excess noise will be $(0.034)^2\times 10$ = 0.012 in %shot noise unit.

Assuming we have a long sequence of pulse quadrature values $X_n$
measured by a BHD, if we consider $X_n$ is contributed by the
($n-1$)th, $n$th, $(n+1)$th pulses, we can write the pulse
quadrature to be $X_n = W_{n}+aW_{n-1}+aW_{n+1}$ ($a$ is a small
number). The correlation coefficient (CC) between consecutive pulses
is CC = $\frac{{E(X_n X_{n+1})}-E(X_n
)E(X_{n+1})}{\sqrt{E(X_n^2)-E^2(X_n)}\sqrt{E(X_{n+1}^2)-E^2(X_{n+1})}}$.
If we assume $E(W_n)=0, E(W_n W_{n+2}) = 0$ (only consecutive pulse
value has a non-zero expectation), CC = $a$. In GMCS QKD, the
variance of Bob's measurement of individual pulse will be
contributed by the variances of its adjacent pulses. With the
quadrature modulation of the coherent state prepared by Alice $V$,
the excess noise due to the overlap between pulses
$\varepsilon_{overlap}$ referring to the input will be $V \times \rm
{CC}^2$.
\end{thebibliography}
\end{document}